# Low Temperature Deposition of Functional Thin Films on Insulating Substrates: Selective Ion Acceleration using Synchronized Floating Potential HiPIMS

Jyotish Patidar[1], Oleksandr Pshyk[1], Lars Sommerhäuser[1], Sebastian Siol[1*]

[1] Empa, Swiss Federal Laboratories for Materials Science and Technology, Dübendorf, Switzerland

*Corresponding author:*
Sebastian Siol, Sebastian.Siol@empa.ch

**Abstract**

Ionized physical vapor deposition techniques, such as high-power impulse magnetron sputtering (HiPIMS) are gaining popularity due to their ability to produce high quality thin films at low deposition temperatures. In those techniques, ions are commonly accelerated onto the growing film using negative potential applied to the substrate. One key challenge however is, how such potentials can be applied on insulating or electrically floating substrates. In this work, we present a novel approach for low-temperature deposition of high-quality thin films on insulating substrates using Synchronized Floating Potential High-Power Impulse Magnetron Sputtering (SFP-HiPIMS). This technique leverages the negative floating potential, induced on the substrate during the HiPIMS discharge. By synchronizing the ion arrival with the substrate's floating potential, specific ions can be accelerated preferentially, thereby enhancing adatom mobility and improving film quality while mitigating the detrimental effects of $Ar^+$ ion bombardment. Our proof-of-concept study demonstrates the deposition of high-quality, textured $Al_{0.8}Sc_{0.2}N$ thin films on various insulating substrates at low temperatures. We show that synchronizing the Al and Sc ion fluxes with the induced negative floating potential significantly enhances the films' crystallinity, c-axis texture and at the same time reduces residual stress. In addition, it enables epitaxial growth on sapphire at temperatures as low as 100°C. The results of this study demonstrate that SFP-HiPIMS provides a practical and economical solution for a long-standing challenge in physical vapor deposition, which can be implemented in standard deposition equipment. SFP-HiPIMS therefore paves the way for advanced manufacturing processes in various emerging technologies.

# 1. Introduction

In recent years, Ionized physical vapor deposition (IPVD) techniques, such as high-power impulse magnetron sputtering (HiPIMS), have garnered significant attention due to their ability to enhance the quality and performance of thin films in various applications.[1], [2] In HiPIMS, power is delivered to the sputter target in short pulses, enabling higher peak power and current densities. The resulting increased plasma densities, in turn, facilitate ionization of the sputtered species. Depending on the deposition parameters ionization rates of over 50 % can be achieved.[2] This increased ionized flux fraction offers unique opportunities for process design, since ions can be accelerated or deflected using electric and/or magnetic fields. Applying a negative potential to the substrates increases the kinetic energy of ions, which enhances adatom mobility through momentum transfer to the growing film. Consequently, ion bombardment during the film growth facilitates the deposition of dense films with improved crystalline properties, even at low growth temperatures.[3], [4] Moreover, ion deflection through substrate biasing has been successfully used for decades in applications such as hard coatings[1] or metallization of vias and trenches in semiconductor device fabrication.[5]

While ion bombardment through biasing offers significant advantages, one downside of using a negative bias on the substrate is that it also attracts abundant process gas ions, such as $Ar^+$, from the plasma. These $Ar^+$ ions can be implanted at interstitial sites within the lattice, leading to the formation of point defects and inducing undesirable compressive stress in the system.[6], [7] Processes like metal-ion synchronized HiPIMS (MIS-HiPIMS) can mitigate the detrimental effects of Ar[+] ion bombardment by selectively only accelerating the film-forming metal ions.[8], [9] This method leverages the different time of flight (ToF) of ions following the HiPIMS discharge. During HiPIMS, for each pulse, the process gas ions tend to arrive earlier at the substrate than the corresponding metal ions due to gas rarefaction. By tailoring the substrate bias potential based on the time-of-flight information, specific ions can be preferentially accelerated while avoiding process gas incorporation into the growing film. This innovative concept has gained a lot of popularity in recent years, as it enables the low temperature deposition of high quality ceramic thin films with minimal compressive stress and structural defects.[8] Most importantly, it holds the promise to use HiPIMS deposition techniques for more defect-sensitive applications, like the development of optical coatings or even semiconducting thin films. We have recently demonstrated that MIS-HiPIMS can produce high-quality and compact AlN and AlScN films at comparably low deposition temperatures, which exhibit a piezoelectric response comparable to that achieved with commercial state-of-the-art processes.[10]

Despite the apparent promise of synchronized ion-acceleration during HiPIMS, one long-standing challenge is holding back its potential use for many applications. While synchronized HiPIMS approaches work well for conductive substrates; their application on insulating or electronically floating substrates is challenging due to the inability to apply electric potentials. This also applies to the growth of thick dielectric films, which depending on their thickness show a similar behavior as insulating substrates.[11] A process capable of achieving this feat would be valuable for future technologies as the demand for high-quality and low-temperature film deposition on insulating substrates is continually increasing. The forthcoming IoT and AI revolution will necessitate a higher integration, but also deposition on temperature sensitive substrates and device stacks for applications such as microelectromechanical systems (MEMS), energy conversion and storage, robotics or biomedical devices.[12], [13], [14], [15], [16], [17] The state-of-the-art method for accelerating ions on an insulating substrate involves applying a radio frequency (RF) plasma to the substrate, thereby inducing a negative floating potential. The surface of an electronically floating or insulating substrate, which is in contact with a plasma will charge negatively, due to the relatively higher mobility of electrons compared to the ions. In the context of RF plasmas, this phenomenon is called self-biasing.[18], [19] RF biasing of insulating substrates, though effective, is limited to higher voltages, which can lead to a substantial amount of ion implantation during film growth, resulting in strain and undesirable defects.[20], [21] In addition, the synchronization of RF bias pulses requires additional hardware (e.g. custom power supplies and synchronization units[22]) and is limited to longer pulses, which also reduces its effectiveness as compared to the synchronized bias voltages used in MIS-HiPIMS. Another commonly reported approach in the literature is bipolar-HiPIMS, where a positive pulse is applied to the target to raise the plasma potential, generating a potential gradient towards the substrate and thus accelerating the sputtered ions. [23], [24] While this approach works well for grounded substrates, it is less effective for insulating or electrically floating substrates. For these cases, the substrate's floating potential often rises with positive pulse preventing an effective acceleration of the ions onto the substrate. Consequently, substantially higher voltages are required to observe any significant effects on the films' growth.[11], [25] To our knowledge, no practical solution exists yet to preferentially accelerate specific ionic species on insulating substrates during pulsed sputter processes like HiPIMS.

In this work, we demonstrate a novel approach for pulsed magnetron sputtering, Synchronized Floating Potential (SFP-HiPIMS), which utilizes the substrate's floating potential to accelerate ions onto the growing film. The negative floating potential observed during an RF-biasing of the substrate can also occur during sputtering (e.g. DC, RF or HiPIMS) when the sputter plasma from an unbalanced magnetron hits the substrate. During the HiPIMS discharge, a negative floating potential is induced on the substrate surface (i.e. when the voltage pulse is applied to the magnetron). This negative potential occurs almost immediately when the sputter pulse is applied and can be used to accelerate positively charged ions onto the substrate. Based on the information of ToF of ionic species, the pulses can be timed in such a way that the floating potential generated by one pulse accelerates the metal ions generated from another pulse. In a proof-of-principle study, we demonstrate the deposition of high-quality, textured AlScN thin films on different insulating or electrically floating substrates at low deposition temperatures. In this study, the Sc-ions originating from one magnetron are accelerated onto the growing film by means of a negative floating potential, which is induced on the substrate by an appropriately synchronized Al discharge from another magnetron. This new deposition scheme can be implemented on standard deposition equipment and is broadly applicable to many materials and substrates. It therefore holds the potential to facilitate the sustainable low-temperature deposition of high-quality thin films for many applications and emerging technologies.

## 2. Results and Discussion

### 2.1 Synchronized Floating Potential HiPIMS: SFP-HiPIMS

During sputtering from an unbalanced magnetron, the substrate is exposed to the sputter plasma. Consequently, due to the higher mobility of electrons, compared to ions, the substrate is negatively charged. If the substrate is insulating or not sufficiently grounded, charges accumulate on the substrate's surface until the floating potential $U_F$ is reached. This phenomenon is commonly referred to as self-biasing.[18] $U_F$ can vary based on the sputter parameters and/or deposition conditions, e.g. the plasma-density or electron energy. SFP-HiPIMS works on the principle of using the negative floating potential $U_F$ to accelerate specific sputtered ions.

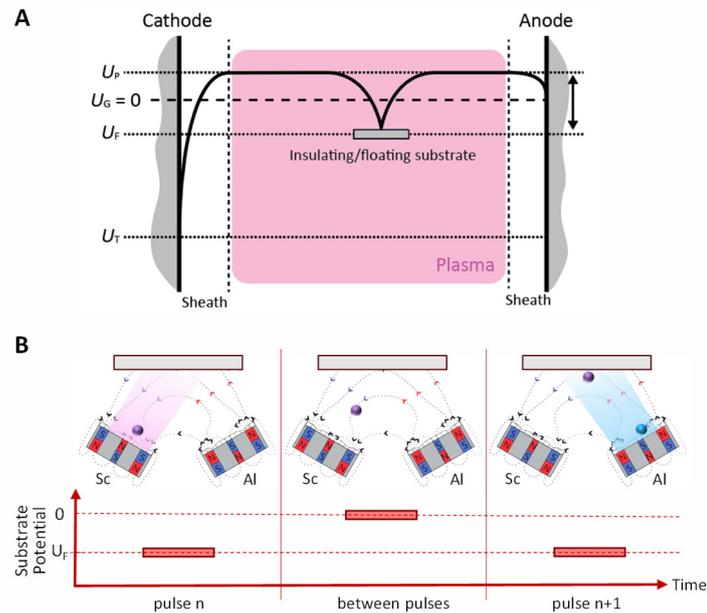

**Figure1: Floating substrate potential utilized to accelerate ions. A)** Schematic illustration of a potential distribution in a sputter chamber showing the floating potential on an insulating substrate ($U_F$), the plasma potential ($U_P$) as well as the chamber and target potentials ($U_G$ and $U_T$, respectively). (Adapted from [18]). **B)** If two magnetrons are used, the ions generated by a HiPIMS pulse on one magnetron (purple) can be accelerated by the floating substrate potential induced by a later HiPIMS pulse on another magnetron (blue). In between the pulses, the floating potential goes to the ground when there is no plasma.

**Figure 1A** represents the potential distribution in a sputter chamber for a single magnetron. Shown are the cathode (i.e. the sputter target), the electrically floating substrate and an anode. Here the anode represents all grounded chamber parts. If the floating substrate is in contact with the sputter plasma it assumes a negative potential, $U_F$, due to the high mobility of electrons. Meanwhile, the plasma, a quasi-neutral entity, maintains a slightly positive potential, $U_P$, thereby ensuring plasma stability. This results in a potential difference $U_P$–$U_F$ in the sheath regions where the acceleration of ions occurs. It is noteworthy that **Figure 1A** is a 2-dimensional representation of potential distribution for a single magnetron, while **Figure 1B** schematically represents the SFP-HiPIMS approach for two magnetrons.

During HiPIMS, the discharge at the magnetron is pulsed, consequently, the substrate is immersed in the sputter plasma predominantly during the sputter pulses. This plasma is formed instantaneously as the power to the target is applied and leads to the generation of a negative floating potential $U_F$ during this HiPIMS discharge, but not necessarily in between pulses. SFP-HiPIMS makes use of this temporally restricted negative floating potential to accelerate ions when they arrive at the substrate. For preferential acceleration of metal-ions, the sputter pulses must be synchronized in a way that a negative floating potential is induced shortly before or in the time frame when the metal-ions arrive at the substrate.

The synchronization condition can be described by means of the delay Δ*t* between the start of two sputter pulses, and the time of flight (ToF) of the respective ionic species. The ToF is measured from the beginning of the HiPIMS pulse to the time the ion flux density on the substrate reaches its maximum. It is proposed that for an effective acceleration of the ions from one pulse using the floating potential induced by the next, the delay Δ*t* should be chosen so that the temporal overlap of the incident ion flux and the induced substrate floating potential is maximized. In addition, the synchronization can be adjusted to specifically avoid the acceleration of undesirable ionic species, such as $Ar^+$ or other process gas ions.

Several pulse configuration setups can be used for the acceleration of ions, including single or multiple magnetrons. In this study, we demonstrate the feasibility of the SFP-HiPIMS approach for the deposition of crystalline AlScN films at low temperatures. For this, we use a confocal setup with multiple magnetrons. In this configuration, we accelerate the Sc ions due to their larger mass through the floating potential generated by Al pulse (see **Figure 1B**). Analogous to the deposition from multiple magnetrons, multiple pulses in the form of a pulse-train can be applied to a single magnetron. Each pulse would generate a burst of ions, which can then be accelerated by the floating potential induced by the following pulse. This way, all ions, but the ones originating from the last pulse in the pulse-train can be accelerated. In an even simpler setup of equipment where pulse-train configuration is not possible, the method would also work for a single magnetron where the frequency can be chosen in such a way that the ToF of ions to be accelerated matches with the time between the HiPIMS pulses, in some instances such conditions could be achieved using pulsed DC sputtering.

Since the method relies on ion acceleration via the generated floating potential of the substrate, its primary limitation is the magnitude of this potential. Typically, this bias is in the range of a few tens of volts, which restricts the kinetic energy of the ions to less than 50 eV. This energy range is ideally suited to significantly enhance adatom mobility without risking detrimental effects due to high energy ion bombardment. Overall, this approach is most effective for depositing thin film materials like AlScN with low lattice displacement thresholds and sensitivity towards point defects. [26], [27], [28] Further advancements to increase the floating potential (e.g. by means of an electron gun) could significantly broaden its applicability for an even wider range of materials.

## 2.2 Proof of concept: Low temperature epitaxial growth of $Al_{0.8}Sc_{0.2}N$

We demonstrate the feasibility of the concept of SFP-HiPIMS in a proof-of-principle experiment. In this experiment, Aluminum Scandium Nitride thin films with 20 at.% Sc ($Al_{0.8}Sc_{0.2}N$) are deposited on insulating or electrically floating substrates using reactive HiPIMS from two Aluminum (Al) and one Scandium (Sc) magnetrons in reactive Ar/N atmosphere in a UHV deposition chamber. AlScN is a piezoelectric and ferroelectric material, which is commonly used for RF-MEMS applications.[29], [30], [31] The material has been intensively investigated in recent years due to its potential for RF filters in 5G and 6G telecommunication technology.[16], [32] We have recently demonstrated that synchronized HiPIMS processes facilitate the growth of high quality AlScN thin films, even at low substrate temperatures and acceleration voltages.[10], [29] However, to date, none of these approaches have been demonstrated on insulating or electronically floating substrates.

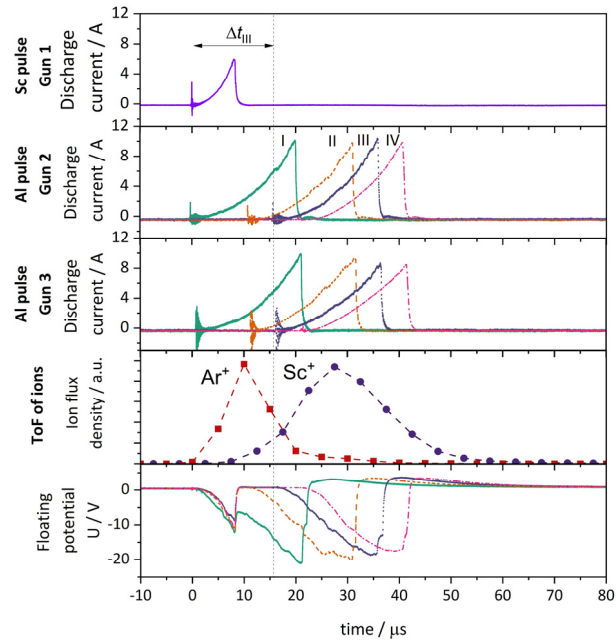

**Figure 2: SFP-HiPIMS deposition of AlScN thin films.** The figure shows the discharge currents of the Sc and Al pulses, along with the time-of-flight of Ar$^+$ and Sc$^+$ ions and the substrate floating potential for four different synchronization modes: **I** : no delay, **II** : $\Delta t_{II}$ =11 μs, **III** : $\Delta t_{III}$ =16 μs, **IV**: $\Delta t_{IV}$ =21 μs. The Sc deposition is operated from one magnetron, while the Al deposition is operated from two magnetrons.

In the proof-of-principle experiment, all three magnetrons are operated in HiPIMS mode. The pulse patterns for all these magnetrons are shown in **Figure 2**, along with the respective generated floating potential and ToF of relevant ions. In the usual mode of operation, all discharges are initiated simultaneously (Mode I). For SFP-HiPIMS synchronization, a delay between the Sc and Al pulses is introduced with the goal to accelerate the incident Sc ions by means of the substrate's floating potential induced during the Al discharges. Figure 2 shows the discharge currents of the Sc and Al pulses, the measured time-of-flight of Ar$^+$ and Sc$^+$ ions and the substrate's floating potential for four different synchronization modes: **I** – no delay, **II** – $\Delta t$ =11 μs, **III** – $\Delta t$ =16 μs, **IV** – $\Delta t$ =21 μs. The delay times were chosen to accelerate the film-forming Sc$^+$ ions, but not the Ar$^+$ process gas ions. Herein, we consider **III** the most promising synchronization scheme, since here the negative floating potential is well aligned with the arrival of Sc$^+$ ions at the substrate. The maximum floating potential reaches around -20 V for the chosen deposition parameters. The magnitude of this potential can be tuned if necessary, by changing the deposition chamber geometry such as working distance, angle of magnetrons, and process conditions such as the sputter power or working pressure.

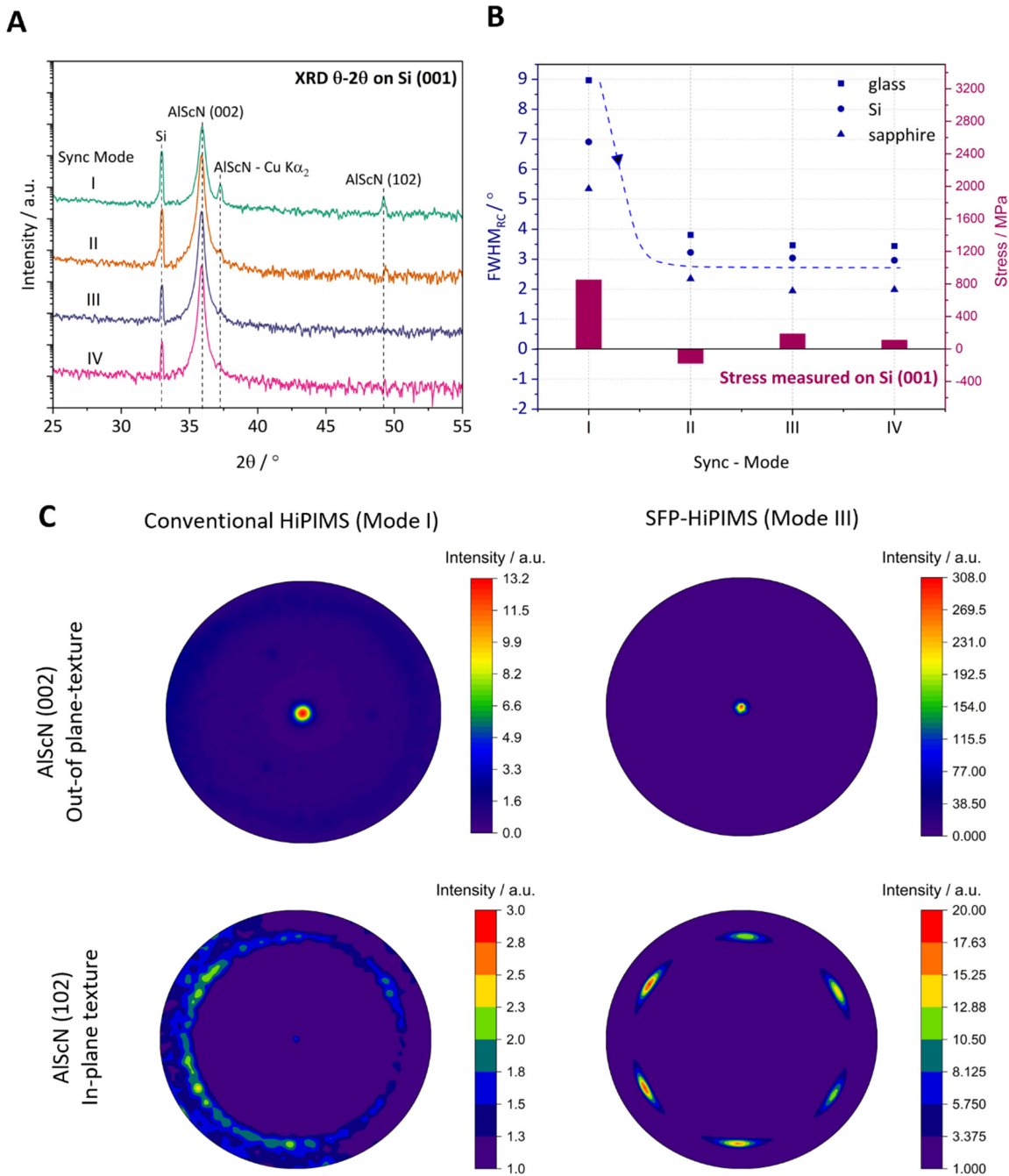

**Figure 3:** Structural analysis and residual stress measurement for AlScN deposited on electrically floating substrates using different synchronization modes. **A)** XRD measurements of AlScN films on Si (001) **B)** FWHM of the AlScN (002) rocking curve for different substrates and residual stress on Si (001) **C)** Pole figure analyses of AlScN deposited on sapphire (001). It is apparent, that the synchronized floating potential significantly improves the crystallinity and texture, but also reduces the tensile residual stress in the films for all tested substrates.

A detailed materials characterization shows significant improvements in the crystalline quality and texture of the films in the case of the SFP-HiPIMS deposition modes (**II-IV**), when compared to the conventional mode of operation (**I**). **Figure 3** shows measurements of the structure as well as residual stress for AlScN films for the four different synchronization modes. In addition, pole figures are added for textural analysis of the AlScN films grown on sapphire (001). The results show an increase in crystallinity and texture for the SFP-HiPIMS depositions, which is indicated by the strongly reduced Full Width at Half Maximum (FWHM) of the (002) rocking curve as well as the disappearing of the AlScN (102) peak in the θ-2θ measurements. As expected, the energetic Sc-ion bombardment leads to a reduction in tensile residual stress from 850 MPa to an almost stress-free state. Most notably, the SFP-HiPIMS deposition mode enables hetero-epitaxial growth of AlScN on sapphire (001) at remarkably low temperatures of 100 °C, while the deposition without synchronization leads to random in-plane orientation (**Figure 3C**).

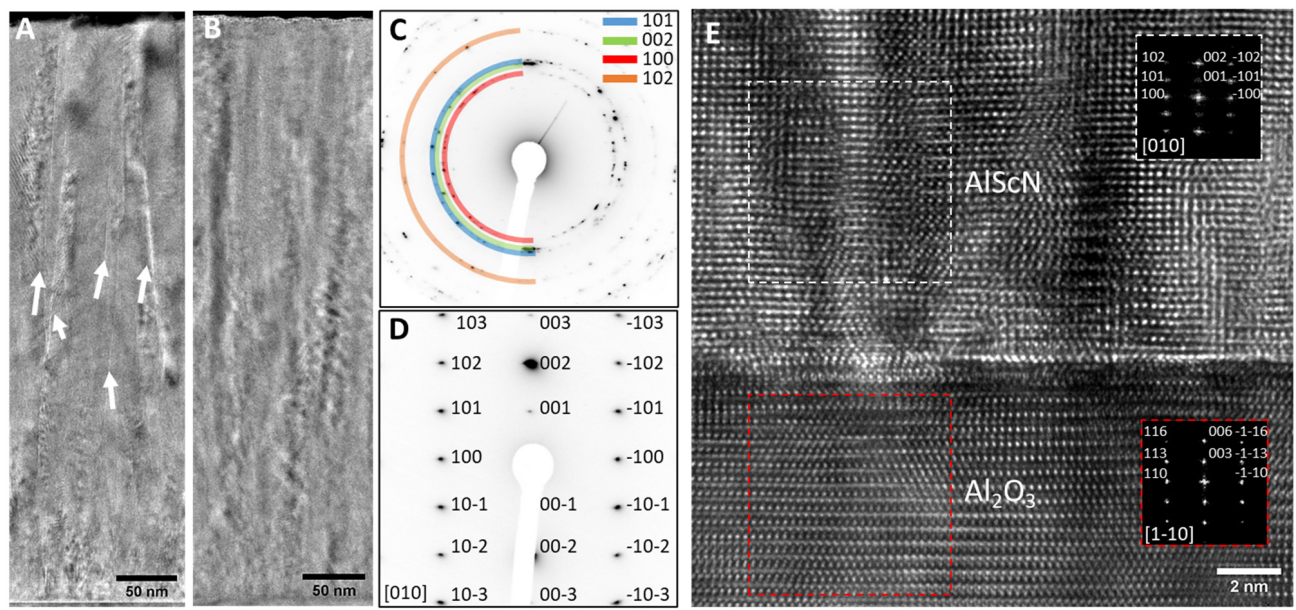

**Figure 4: Microstructure and epitaxial growth**. Low-magnification BF-TEM cross-sectional images from AlScN thin films grown on the sapphire (001) substrate using sync mode I with arrows pointing on the inter-columnar voids **(A)** and SFP-HiPIMS mode III resulting in a compact microstructure **(B)**. SAED patterns from AlScN thin films grown on sapphire (001) substrates using mode I **(C)** and SFP-HiPIMS mode III **(D)**. HR-TEM image acquired at the film/substrate interface for the film grown using SFP-HiPIMS mode III with fast Fourier transforms (FFT) patterns as the inserts acquired from the film and the substrate.

The structural analysis is further confirmed by cross-sectional transmission electron microscopy (TEM) images (**Figure 4**). Bright field TEM images reveal that AlScN thin films grown in mode I have inter-columnar voids visible as bright areas extended along the growth direction (**Fig. 4A**), whereas the films grown using SFP-HiPIMS mode III have dense microstructure with no open grain boundaries (**Fig. 4B**). This is in agreement with the reduction of tensile residual stresses. The observed densification is a direct consequence of an ion-induced densification process, which becomes more efficient when the plasma-induced floating potential is synchronized with the Sc$^+$-rich portion of the pulse. Moreover, selected area electron diffraction (SAED) patterns reveal that the film grown in mode I is polycrystalline (**Fig. 4C**) whereas the one grown in SFP-HiPIMS mode III shows (002) out-of-plane texture and a diffraction pattern typical for single crystals (**Fig. 4D**). A high-resolution TEM (HR-TEM) image acquired from the film/substrate interface of the film grown in SFP-HiPIMS mode III (**Fig. 4E**) demonstrates hetero-epitaxy of the AlScN film on sapphire (001) with an out-of-plane AlScN [001]//Al$_2$O$_3$[001] and in-plane AlScN [100]//Al$_2$O$_3$[110] epitaxial relationship.

These results highlight the utility of SFP-HiPIMS for the deposition of highly-textured thin film materials at low deposition temperatures. This is particularly interesting for the deposition of functional thin films on temperature sensitive substrates or device stacks. But the selective metal-ion acceleration enabled by SFP-HiPIMS is also useful for other applications. Analogous to the proof-of-principle experiment outlined in Figure 2, HiPIMS depositions from two Ta targets were performed in Ar/N atmosphere. By synchronizing one Ta pulse to the ToF of Ta$^+$ ions from the respective preceding Ta pulse, the hardness of the cubic TaN films was improved by almost 20%. (See Supporting Information S1). This demonstrates the universal applicability of the concept.

## 3. Conclusion

The investigation reports a novel approach for preferentially accelerating specific ions during thin film deposition on insulating or electrically floating substrates – synchronized floating potential HiPIMS, or SFP-HiPIMS. The approach works on the principle of utilizing the negative floating potential generated during the HiPIMS pulse when the substrate is immersed in the sputter plasma. With appropriate synchronization, this potential can be used to accelerate specific ions onto the substrate. For the proof of concept, we investigated the effect of SFP-HiPIMS on the properties of AlScN films grown on a variety of insulating or electrically floating substrates. The setup involved sputtering from three magnetrons with two Al and one Sc target in a reactive environment and a confocal geometry. The Sc ions were chosen to be accelerated due to their higher ToF and heavier mass, thus imparting more momentum through ion bombardment. The novel synchronization scheme leads to a significant improvement in the film quality. We were able to show that the ion bombardment with low kinetic energies around the lattice displacement threshold significantly improves the crystallinity, texture and stress state. For sapphire (001) substrates, it even facilitates epitaxial growth at low deposition temperatures, highlighting the potential of SFP-HiPIMS for the growth of high-quality thin films on insulating substrates at low substrate temperatures.

Overall, SFP-HiPIMS provides a practical and low-cost solution for the tailored acceleration of ions onto insulating substrates, a fundamental challenge in ionized physical vapor deposition (IPVD) which had not been solved to date. In addition, the new deposition scheme holds the promise of extending the use cases of IPVD techniques, such as HiPIMS, to more advanced applications with the goal of improving film quality, lowering the overall processing temperatures, but also enabling unique or improved functionalities. This is particularly relevant considering the increasing demands for sustainable manufacturing processes and increasingly temperature-sensitive device structures in emerging electronic applications. But also other technologies benefit from the approach. The concept can be implemented for a large number of materials and substrates and therefore can benefit various applications and technologies. In addition, it does not require any specialized equipment and can be implemented with little additional cost on most commercial sputter systems.

## 4. Experimental Section

The AlScN films were deposited using a custom-built by AJA ATC 1800 chamber, with a base pressure of <$10^{-6}$ Pa. The films were deposited on multiple substrates including p-type Si (001), Eagle-XG glass substrate and sapphire (001) substrate (CrysTec AG). The type-2 unbalanced magnetrons in the chamber are equipped with 2 inch Al (Lesker, purity: 99.999 at.%) and Sc targets (Hunan Advanced Metal Materials, 99.999 at %, O < 800 ppm) in confocal geometry. The substrates were put at a working distance of 12 cm and were heated from the back using 5 halogen lamps, ensuring uniform heat distribution across the whole sample holder. All the depositions were performed at a pressure of 5 µbar, with the flow rates of Ar and $N_2$ in the chamber maintained at 20/12 (sccm/sccm). The power to the sputtering target was provided using HiPSTER 1 power supply by Ionautics and the pulses were synced in time using an Ionautics synchronization unit. The discharge currents, voltages and consequent floating potential for the HiPIMS pulses are recorded using a Tektronix oscilloscope with multiple channels, attached to the HiPIMS power supplies and substrate holder. The depositions were performed at 100°C to maintain a constant, low temperature throughout the deposition. The substrates were cleaned in an ultrasonicator in acetone and ethanol individually, and further soaked at 300°C in the deposition chamber.

For the synchronization of the substrate's floating potential, the ToF of the Al and Sc ions was first estimated by time-resolved mass spectrometry measurement using a Hiden Analytical EQP-300. The orifice of the spectrometer, 50 µm in diameter, was grounded and placed at the working distance while facing the magnetron. The triggering signal was provided by the pulsing unit of the HiPIMS power supply attached to the target. The gate width was set to 5 µs, consistent with the step size of the measurement. The time-resolved measurements were performed for the relevant ionic species such as $^{36}$Ar and $^{45}$Sc. A less abundant isotope of Ar is used here to avoid the saturation of the detector. The time-of-flight in the mass spectrometer is calibrated by applying a gating potential at the driven front end of the spectrometer. A more detailed description of this method can be found in our previous publication.[10]

X-ray diffraction (XRD) analysis of the films was performed using a Bruker D8 in Bragg-Brentano geometry and Cu-kα radiation. The pole figures were obtained for the films for psi ranging from 0° to 80° with a step size of 3° in psi and phi. The thickness of the films was measured using a Bruker Dektak XT profilometer. The cation concentration was determined using X-ray fluorescence in a commercial system featuring Rh irradiation (Helmut Fischer SDV-SDD). The optics were flushed with He during the measurement to increase the sensitivity for light elements (i.e. Al).

AFM images were acquired using Bruker Nanoscope in scan-asyst mode using Si cantilevers with spring constant of 0.4 N/m. The images are further processed in Gwyddion. Nanoindentation measurements are performed using a Picodentor HM500 equipped with a Berkovich diamond tip. The indentation depth is limited to less than 15 % of the total film thickness to minimize the elastic behavior from the substrate during nanoindentation. 30 indents are performed for each sample to determine the average value and standard deviation. Hardness was determined following Oliver and Pharr method.

A JEOL LEM-22000FS microscope operated at 200 kV was used for transmission electron microscopy analysis. TEM specimens were prepared by focused ion beam (FIB) lift-out procedure in a FEI Helios NanoLab G3 UC Dual Beam SEM/Ga$^+$ FIB system.

# References

[1] A. Anders, "Tutorial: Reactive high power impulse magnetron sputtering (R-HiPIMS)," *J. Appl. Phys.*, vol. 121, no. 17, 2017, doi: 10.1063/1.4978350.

[2] U. Helmersson, M. Lattemann, J. Bohlmark, A. P. Ehiasarian, and J. T. Gudmundsson, "Ionized physical vapor deposition (IPVD): A review of technology and applications," *Thin Solid Films*, vol. 513, no. 1–2, pp. 1–24, 2006, doi: 10.1016/j.tsf.2006.03.033.

[3] F. Cemin *et al.*, "Epitaxial growth of Cu(001) thin films onto Si(001) using a single-step HiPIMS process," *Sci. Rep.*, vol. 7, no. 1, p. 1655, May 2017, doi: 10.1038/s41598-017-01755-8.

[4] X. Bai, Q. Cai, W. Xie, Y. Zeng, C. Chu, and X. Zhang, "In-situ crystalline TiNi thin films deposited by HiPIMS at a low substrate temperature," *Surf. Coatings Technol.*, vol. 455, p. 129196, Feb. 2023, doi: 10.1016/j.surfcoat.2022.129196.

[5] L. Jablonka, P. Moskovkin, Z. Zhang, S.-L. Zhang, S. Lucas, and T. Kubart, "Metal filling by high power impulse magnetron sputtering," *J. Phys. D. Appl. Phys.*, vol. 52, no. 36, p. 365202, Sep. 2019, doi: 10.1088/1361-6463/ab28e2.

[6] G. Greczynski *et al.*, "Selection of metal ion irradiation for controlling Ti 1-xAl xN alloy growth via hybrid HIPIMS/magnetron co-sputtering," *Vacuum*, vol. 86, no. 8, pp. 1036–1040, 2012, doi: 10.1016/j.vacuum.2011.10.027.

[7] T. Shimizu *et al.*, "Low temperature growth of stress-free single phase α-W films using HiPIMS with synchronized pulsed substrate bias," *J. Appl. Phys.*, vol. 129, no. 15, 2021, doi: 10.1063/5.0042608.

[8] G. Greczynski, I. Petrov, J. E. Greene, and L. Hultman, "Paradigm shift in thin-film growth by magnetron sputtering: From gas-ion to metal-ion irradiation of the growing film," *J. Vac. Sci. Technol. A*, vol. 37, no. 6, p. 060801, 2019, doi: 10.1116/1.5121226.

[9] G. Greczynski *et al.*, "A review of metal-ion-flux-controlled growth of metastable TiAlN by HIPIMS/DCMS co-sputtering," *Surf. Coatings Technol.*, vol. 257, pp. 15–25, 2014, doi: 10.1016/j.surfcoat.2014.01.055.

[10] J. Patidar *et al.*, "Improving the crystallinity and texture of oblique-angle-deposited AlN thin films using reactive synchronized HiPIMS," *Surf. Coatings Technol.*, vol. 468, p. 129719, Sep. 2023, doi: 10.1016/j.surfcoat.2023.129719.

[11] H. Du, M. Zanáška, N. Brenning, and U. Helmersson, "Bipolar HiPIMS: The role of capacitive coupling in achieving ion bombardment during growth of dielectric thin films," *Surf. Coatings Technol.*, vol. 416, p. 127152, Jun. 2021, doi: 10.1016/j.surfcoat.2021.127152.

[12] K.-K. Liu *et al.*, "Growth of Large-Area and Highly Crystalline MoS 2 Thin Layers on Insulating Substrates," *Nano Lett.*, vol. 12, no. 3, pp. 1538–1544, Mar. 2012, doi: 10.1021/nl2043612.

[13] M. Hassan *et al.*, "Significance of Flexible Substrates for Wearable and Implantable Devices: Recent Advances and Perspectives," *Adv. Mater. Technol.*, vol. 7, no. 3, Mar. 2022, doi: 10.1002/admt.202100773.

[14] B. P. Nabar, Z. Celik-Butler, and D. P. Butler, "Self-Powered Tactile Pressure Sensors Using Ordered Crystalline ZnO Nanorods on Flexible Substrates Toward Robotic Skin and Garments," *IEEE Sens. J.*, vol. 15, no. 1, pp. 63–70, Jan. 2015, doi: 10.1109/JSEN.2014.2337115.

[15] R.-H. Kim *et al.*, "Waterproof AlInGaP optoelectronics on stretchable substrates with applications in biomedicine and robotics," *Nat. Mater.*, vol. 9, no. 11, pp. 929–937, Nov. 2010, doi: 10.1038/nmat2879.

[16] G. Giribaldi, L. Colombo, P. Simeoni, and M. Rinaldi, "Compact and wideband nanoacoustic pass-band filters for future 5G and 6G cellular radios," *Nat. Commun. 2024 151*, vol. 15, no. 1, pp. 1–13, Jan. 2024, doi: 10.1038/s41467-023-44038-9.

[17] B. Luo *et al.*, "Magnetoelectric microelectromechanical and nanoelectromechanical systems for the IoT," *Nat. Rev. Electr. Eng. 2024 15*, vol. 1, no. 5, pp. 317–334, May 2024, doi: 10.1038/s44287-024-00044-7.

[18]   P. M. Martin, "Handbook of deposition technologies for films and coatings: science, applications and technology," p. 912, 2010.

[19]   *Principles of Vapor Deposition of Thin Films*. Elsevier, 2006. doi: 10.1016/B978-0-08-044699-8.X5000-1.

[20]   B. Abdallah, A. Chala, P.-Y. Jouan, M. P. Besland, and M. A. Djouadi, "Deposition of AlN films by reactive sputtering: Effect of radiofrequency substrate bias," *Thin Solid Films*, vol. 515, no. 18, pp. 7105–7108, Jun. 2007, doi: 10.1016/j.tsf.2007.03.006.

[21]   N. H. Azhan, K. Su, K. Okimura, and J. Sakai, "Radio frequency substrate biasing effects on the insulator-metal transition behavior of reactively sputtered VO2 films on sapphire (001)," *J. Appl. Phys.*, vol. 117, no. 18, May 2015, doi: 10.1063/1.4921105.

[22]   V. BABAYAN, Z. Q. HUA, M. WU, A. M. ALLEN, and B. CITLA, "Sync controller for high impulse magnetron sputtering," *US Pat.*, Art. no. US 2019 / 0088457 A1, 2019.

[23]   V. Tiron *et al.*, "Overcoming the insulating materials limitation in HiPIMS: Ion-assisted deposition of DLC coatings using bipolar HiPIMS," *Appl. Surf. Sci.*, vol. 494, pp. 871–879, Nov. 2019, doi: 10.1016/j.apsusc.2019.07.239.

[24]   R. P. B. Viloan, J. Gu, R. Boyd, J. Keraudy, L. Li, and U. Helmersson, "Bipolar high power impulse magnetron sputtering for energetic ion bombardment during TiN thin film growth without the use of a substrate bias," *Thin Solid Films*, vol. 688, p. 137350, Oct. 2019, doi: 10.1016/j.tsf.2019.05.069.

[25]   M. Zanáška *et al.*, "Dynamics of bipolar HiPIMS discharges by plasma potential probe measurements," *Plasma Sources Sci. Technol.*, vol. 31, no. 2, p. 025007, Feb. 2022, doi: 10.1088/1361-6595/ac4b65.

[26]   A. Y. Konobeyev, U. Fischer, Y. A. Korovin, and S. P. Simakov, "Evaluation of effective threshold displacement energies and other data required for the calculation of advanced atomic displacement cross-sections," *Nucl. Energy Technol.*, vol. 3, no. 3, pp. 169–175, Sep. 2017, doi: 10.1016/j.nucet.2017.08.007.

[27]   S. Anderson, M. Khafizov, and A. Chernatynskiy, "Threshold displacement energies and primary radiation damage in AlN from molecular dynamics simulations," *Nucl. Instruments Methods Phys. Res. Sect. B Beam Interact. with Mater. Atoms*, vol. 547, p. 165228, Feb. 2024, doi: 10.1016/j.nimb.2023.165228.

[28]   A. S. Hauck, M. Jin, and B. R. Tuttle, "Atomic displacement threshold energies and defect generation in GaN, AlN, and AlGaN: A high-throughput molecular dynamics investigation," *Appl. Phys. Lett.*, vol. 124, no. 15, Apr. 2024, doi: 10.1063/5.0190371.

[29]   K.-H. Kim *et al.*, "Scalable CMOS back-end-of-line-compatible AlScN/two-dimensional channel ferroelectric field-effect transistors," *Nat. Nanotechnol.*, vol. 18, no. 9, pp. 1044–1050, Sep. 2023, doi: 10.1038/s41565-023-01399-y.

[30]   D. K. Pradhan *et al.*, "A scalable ferroelectric non-volatile memory operating at 600 °C," *Nat. Electron.*, vol. 7, no. 5, pp. 348–355, Apr. 2024, doi: 10.1038/s41928-024-01148-6.

[31]   Y. Zou *et al.*, "Aluminum scandium nitride thin-film bulk acoustic resonators for 5G wideband applications," *Microsystems Nanoeng.*, vol. 8, no. 1, p. 124, Nov. 2022, doi: 10.1038/s41378-022-00457-0.

[32]   G. Giribaldi *et al.*, "X-Band Multi-Frequency 30% Compound SCALN Microacoustic Resonators and Filters for 5G-Advanced and 6G Applications," in *2022 Joint Conference of the European Frequency and Time Forum and IEEE International Frequency Control Symposium (EFTF/IFCS)*, IEEE, Apr. 2022, pp. 1–4. doi: 10.1109/EFTF/IFCS54560.2022.9850563.

[33]   J. Patidar, K. Thorwarth, T. Schmitz-kempen, R. Kessels, and S. Siol, "Deposition of highly-crystalline AlScN thin films using synchronized HiPIMS – from combinatorial screening to piezoelectric devices TOC Figure," *arxiv*, Apr. 2024, [Online]. Available: http://arxiv.org/abs/2405.00210

*Supporting Information for:*

**Low Temperature Deposition of Functional Thin Films on Insulating Substrates: Selective Ion Acceleration using Synchronized Floating Potential HiPIMS**

Jyotish Patidar[1], Oleksandr Pshyk[1], Lars Sommerhäuser[1], Sebastian Siol[1*]

[1] Empa, Swiss Federal Laboratories for Materials Science and Technology, Dübendorf, Switzerland

*\* Corresponding author:*

*Sebastian Siol, Sebastian.Siol@empa.ch*

Keywords: insulating substrate, floating substrate, synchronized HiPIMS, substrate bias, AlScN, epitaxial growth

# S1: Results on SFP-HiPIMS for deposition of denser TaN films

To verify the applicability of SFP-HiPIMS on other material systems, we also deposited TaN films with conventional synchronization (i.e. all pulses start at same time) and SFP-HiPIMS (synchronized with Ta$^+$ metal ion flux).

A minor increase in measured stress values (using wafer curvature) was observed from -2.26 GPa to -2.52 GPa with SFP-HiPIMS. However, the ion bombardment led to increased hardness from 14.99±0.97 GPa to 17.83±0.92 GPa, which cannot be caused by the marginal increase in the compressive stresses. Therefore, we assign this improvement in mechanical properties to the densification of the film that is in agreement with previous studies on refractory nitride thin films densified by ion irradiation.[1] The change in the films' properties is less prominent here as compared to AlScN, potentially due to higher lattice displacement threshold of Ta as compared to Al or Sc.[2]

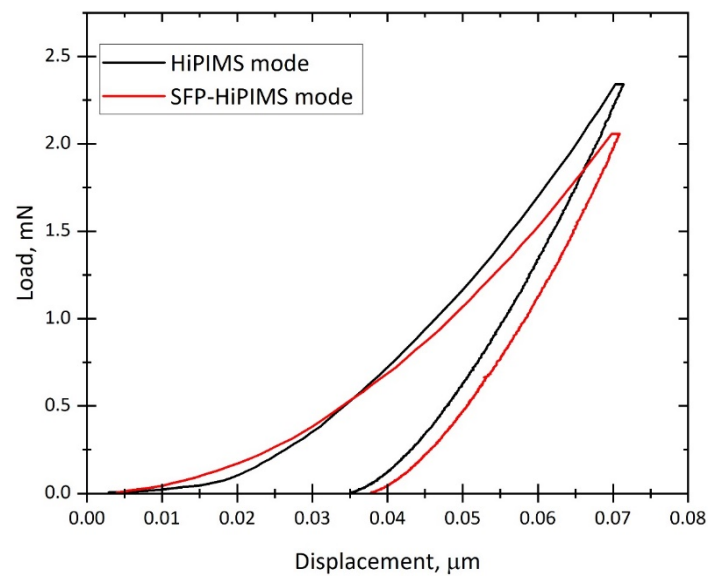

Fig. S1: Load-displacement curves for TaN films deposited with conventional synchronization and SFP-HiPIMS.

## S2: Surface morphology of deposited films

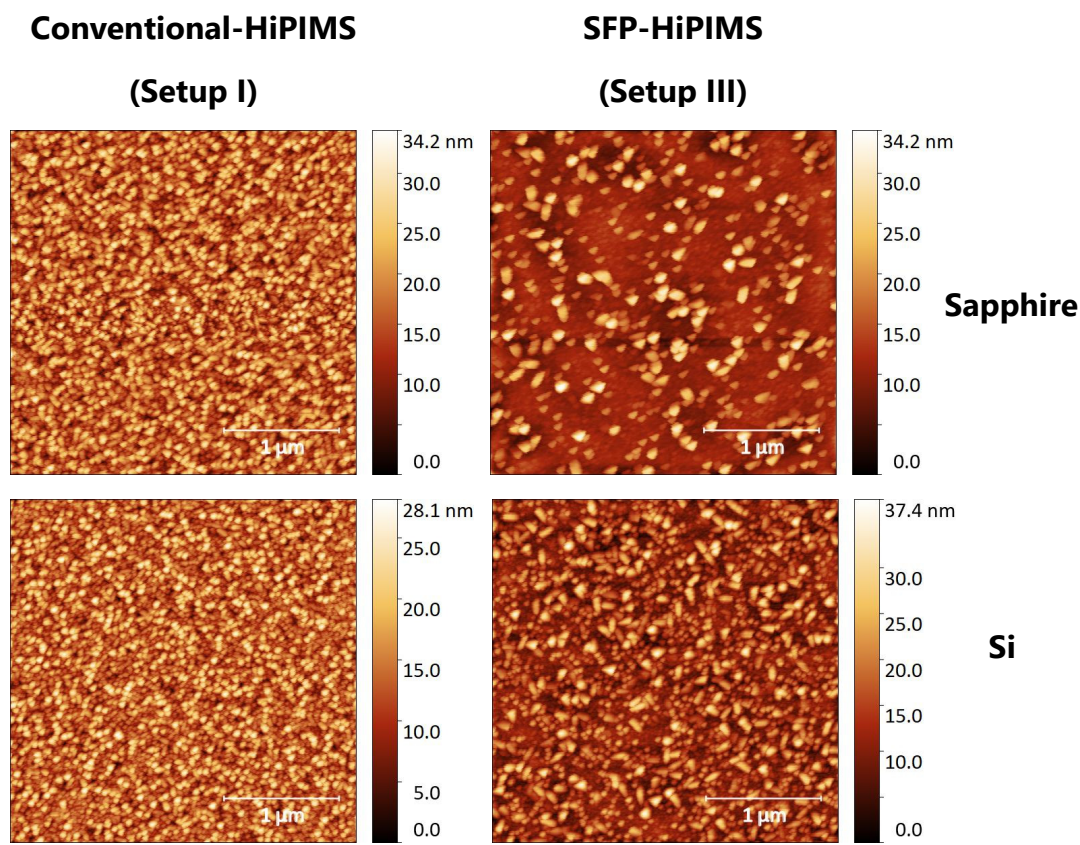

Fig. S2 : AFM images of samples deposited with synchronized setup A and C on sapphire and Si substrate.

The AFM characterization reveals the surface morphology of films deposited using conventional synchronization and SFP-HiPIMS techniques. While the samples have similar thicknesses, the SFP-HiPIMS deposited films display the nucleation of larger grains. This is attributed to enhanced adatom mobility resulting from increased ion bombardment.

# References

[1]  A. V. Pshyk, I. Petrov, B. Bakhit, J. Lu, L. Hultman, and G. Greczynski, "Energy-efficient physical vapor deposition of dense and hard Ti-Al-W-N coatings deposited under industrial conditions," *Mater. Des.*, vol. 227, p. 111753, Mar. 2023, doi: 10.1016/j.matdes.2023.111753.

[2]  A. Y. Konobeyev, U. Fischer, Y. A. Korovin, and S. P. Simakov, "Evaluation of effective threshold displacement energies and other data required for the calculation of advanced atomic displacement cross-sections," *Nucl. Energy Technol.*, vol. 3, no. 3, pp. 169–175, Sep. 2017, doi: 10.1016/j.nucet.2017.08.007.